\documentclass{PoS}
\usepackage{slashed}
\usepackage{mathtools}

\newcommand{\beqn}{\begin{eqnarray}}
\newcommand{\eeqn}{\end{eqnarray}}
\newcommand{\eq}[1]{(\ref{#1})}
\newcommand{\bum}{$\blacktriangleright$\ }
\newcommand{\nm}{\,${\mathrm{nm}}$\,}
\newcommand{\avr}[1]{{\left\langle #1 \right\rangle}}
\newcommand{\Z}{{\mathbb Z}}

\newcommand{\cL}{{\cal L}}
\newcommand{\bx}{\boldsymbol {x}}
\newcommand{\Cas}{{\mathrm{Cas}}}

\title{Nonperturbative Casimir Effects in Field Theories: aspects of confinement, dynamical mass generation and chiral symmetry breaking}

\ShortTitle{Nonperturbative Casimir Effects in Field Theory}

\author{\speaker{M. N. Chernodub}
\\
Institut Denis Poisson UMR 7013, Universit\'e de Tours, 37200 France\\
Laboratory of Physics of Living Matter, Far Eastern Federal University, Sukhanova 8, Vladivostok, 690950, Russia\\
E-mail: \email{maxim.chernodub@idpoisson.fr}}
\author{V. A. Goy\\
Laboratory of Physics of Living Matter, Far Eastern Federal University, Sukhanova 8, Vladivostok, 690950, Russia}
\author{A. V. Molochkov\\
Laboratory of Physics of Living Matter, Far Eastern Federal University, Sukhanova 8, Vladivostok, 690950, Russia}

\abstract{The Casimir effect is a quantum phenomenon rooted in the fact that vacuum fluctuations of quantum fields are affected by the presence of physical objects and boundaries. Since the energy spectrum of the vacuum fluctuations depends on distances between (and geometries of) physical bodies, the quantum vacuum exerts a small but experimentally detectable force on neutral objects. Usually, the associated Casimir energy is calculated for free or weakly coupled quantum fields. We review recent studies of the Casimir effect in field-theoretical models which mimic features of non-perturbative QCD such as chiral or deconfining phase transitions. We discuss ${{\mathbb C}P}^{\,N-1}$ sigma model and chiral Gross-Neveu model in (1+1) dimensions as well as compact U(1) gauge theory and Yang-Mills theory in (2+1) dimensions.}

\FullConference{XIII Quark Confinement and the Hadron Spectrum - Confinement2018\\
		31 July - 6 August 2018\\
		Maynooth University, Ireland}

\begin{document}

\section{Introduction}

The presence of physical (material) objects affects virtual (vacuum) fluctuations nearby. The total energy of vacuum fluctuations depends on shapes, orientations, and distances between physical bodies so that the bodies experience mutual forces related to the minimization of the vacuum energy. This mechanism is a cornerstone of the Casimir effect~\cite{Casimir:1948dh}, which implies that even neutral physical bodies may interact which each other via forces of the vacuum origin. Although the energy density of the vacuum fluctuations is given by a divergent integral and/or sum, the excess in the vacuum energy caused by the presence of the physical bodies is a finite quantity. The difference between the two is called the Casimir energy.

A simplest realization of the Casimir effect is given by two parallel plates made of a perfectly conducting metal. The plates act as boundaries which restrict fluctuations of the electromagnetic field. Since the corresponding Casimir energy per unit plate area (``the Casimir pressure''),
\beqn
\frac{\avr{E}}{\mathrm{Area}} = - \frac{\pi^2}{720} \frac{\hbar c}{R^3},
\label{eq:E:Casimir}
\eeqn
diminishes with the decrease the separation $R$ between the plates, the Casimir effect leads to an attractive force between the plates. Although this quantum force is extremely small at human scales, the Casimir pressure reaches the tremendous value in 1 atmosphere at the ``optimistic separation'' in $R=10$\nm between the plates.

The effect has been observed experimentally~\cite{Lamoreaux:1996wh,Mohideen:1998iz} down to separations of 100\nm between a high\-ly\--conducting metallic sphere and a plate with the accuracy of about 1\% in Ref.~\cite{Mohideen:1998iz} and in further experiments (the double-plate geometry is difficult to realize in practice due to geometrical imperfections of physical plates). The Casimir effect is an important phenomenon not only in view of its possible future applications in technology, but also due to its fundamental value which indirectly demonstrates the physical significance of the vacuum energy. In our short review we leave aside the important theoretical question if the Casimir effect is a ``decisive'' evidence of the existence of the zero-point energies of quantum fields or not~\cite{Jaffe:2005vp}. We treat the Casimir energy as a real observable quantity since the associated force have indeed been observed in the real physical experiments. Excellent reviews of the Casimir effect may be found in Refs.~\cite{ref:Bogdag,ref:Milton}. 

We discuss how (self-) interactions of quantum fields affect the vacuum properties in a finite Casimir geometry and vice versa. In the phenomenologically interesting case of quantum electrodynamics, radiative corrections to the Casimir effect can be calculated in a perturbation theory. Despite of the weakness of the electromagnetic interaction, the fluctuations of photons and electrons in a finite geometry may lead to surprising results such as the Scharnhorst effect~\cite{Scharnhorst:1990sr} (Section~\ref{eq:perturbative}). 

In strongly coupled theories the interactions may not only lead so a substantial modification of the Casimir energy, but they may also affect the nonperturbative structure of the vacuum itself. We discuss the effects of the finite geometry confinement (Section~\ref{eq:confinement}) and dynamical breaking of the chiral symmetry (Section~\ref{eq:chiral}) in certain model-based and first-principles calculations.

We leave aside finite geometry effects which emerge in interacting theories possessing second-order thermal phase transitions: Due to a divergent correlation length, thermal fluctuations exert thermodynamic Casimir-like forces on a boundary of the system~\cite{ref:FdG}. In spin models, the thermodynamic Casimir effect has been studied in details in Refs.~\cite{ref:Hasenbusch}. We also do not touch the powerful world-line numerical approaches to the Casimir effects in nontrivial geometries~\cite{ref:Gies}.

\clearpage

\section{Perturbative corrections to the Casimir effect}
\label{eq:perturbative}

The Casimir phenomenon, in its original formulation, arises due to quantum fluctuations of free fields in a vacuum. The presence of boundaries affects the spectrum of virtual particles (quantum fluctuations), and, expectedly, their vacuum energy. In interacting field theories, the energy spectrum of the virtual particles is affected not only by the boundary of the system but also by the (self-) interaction between the particles. This interaction modifies the Casimir energy. 

Below we will briefly discuss the influence of the perturbative interactions on the Casimir forces in quantum electrodynamics which is a well-understood weakly interacting theory. Then we will proceed to explore certain nonperturbative Casimir effects in strongly interacting theories that exhibit chiral symmetry breaking and confinement phenomena.

\vskip 3mm
\noindent 
{\bf\small{\bum Perturbative correction to the Casimir energy}}. The Casimir effect, despite its quantum nature, is determined by a tree-level physics. In the original proposal of H.~Casimir, the vacuum modes of free virtual photons are modified by two ideal narrowly-placed parallel plates, leading to a drop in the electromagnetic energy-density in between the plates~\eq{eq:E:Casimir}. In quantum electrodynamics, the tree-level Casimir energy is affected by radiative corrections which appear as a result of interactions of the (virtual) photons with (virtual) electrons and positrons. A leading perturbative correction to the tree-level Casimir effect is given by a one-loop process in which a virtual photon scatters off one plate, creates a virtual electron-positron pair that annihilates into another photon which, finally, hits the opposite plate. Due to this process, the Casimir energy acquires the radiative correction coming from photonic and fermionic fluctuations~\cite{Bordag:1983zk}:
\beqn
\frac{\avr{\delta E}}{\mathrm{Area}} = - \frac{\pi^2}{720} \frac{\hbar c}{R^3} \left( 1 - \frac{9 \alpha_{\,\mathrm{e.m.}} \hbar}{32 m_e c} \frac{1}{R} \right)\,, 
\qquad
\alpha_{\,\mathrm{e.m.}} = \frac{e^2}{4 \pi} \approx \frac{1}{137},
\label{eq:E:Casimir:delta}
\eeqn
where $m_e$ is the mass of the electron.
The radiative contribution in Eq.~\eq{eq:E:Casimir:delta} is, expectedly, very small: even at our ``optimistic'' $R=10$\nm separation bet\-ween the plates, it gives a $10^{-7}$ correction to the leading free-photon energy~\eq{eq:E:Casimir}. Needless to say that an observation of this tiny radiative effect is incompatible with the existing experimental technology.

\vskip 3mm
\noindent 
{\bf\small{\bum The Scharnhorst effect}}. Radiative corrections in the double-plate configuration lead also to an unexpected result: it turns out that the velocity of a low-frequency light, propagating in the normal direction with respect to the plates, exceeds the standard velocity of light $c$ by the amount:
\beqn
\delta c = + \frac{11 \pi^2}{90^2} \alpha_{\,\mathrm{e.m.}}^2\left( \frac{\hbar}{m_e c} \frac{1}{R} \right)^4 > 0 \,.
\label{eq:Scharnhorst}
\eeqn
In other words, a photon travels in the space between the plates faster than in an unbounded vacuum outside the plates. This is the Scharnhorst effect~\cite{Scharnhorst:1990sr}. The correction~\eq{eq:Scharnhorst} emerges in the second-order perturbation theory as a light-by-light scattering process which proceeds via a fermion-box loop: a pair of legs corresponds to a travelling photon while another pair of legs is a closed loop of a virtual photon that scatters off the both boundaries. The positive addition to the speed of light~\eq{eq:Scharnhorst}  is too small to be observed in practice as it amounts to a $10^{-24}$ correction to the speed of light at our benchmark distance in $R = 10$\nm. Notice that despite the Scharnhorst effect formally implies a suspicious ``faster-than-light travel'', it cannot be used to create causal paradoxes~\cite{ref:causality}.

\clearpage
\section{Casimir effect, mass gap generation and (de)confinement}
\label{eq:confinement}

Consider now a nonperturbative vacuum of a confining (gauge) theory in which a particle and an antiparticle are attracted to each other by a linear potential at large enough distances. One may think about Quantum Chromodynamics in which the fundamental color degrees of freedom, quarks and gluons, are confined into colorless states, hadrons and glueballs. The nonperturbative confining force is closely related to the dynamical mass generation in Yang-Mills theory.

Let us impose the Casimir conditions on the confining fields. How the confinement and mass gap generation influences the Casimir energy? And, vice versa, what is the effect of the finite Casimir geometry on the confining properties and the phase structure of the theory? Since in QCD these questions are of a nonperturbative nature, it is natural to address them first in appropriate toy models and then in first-principles lattice simulations of Yang-Mills theory.

Before proceeding further we would like to notice that a finite Casimir geometry is not always equivalent to a finite (closed) volume. In a finite volume a smallest nonzero momentum increases as the volume shrinks to zero. This momentum serves as an infrared cutoff of the theory, which -- in a sufficiently small volume -- may become higher than the ultraviolet (perturbative) scale of the theory. As a result, asymptotically free theories in a shrinking volume become weakly coupled and gradually lose their non-perturbative properties (for example, Yang-Mills theories in a finite volume become deconfining at zero temperature). In addition, in a finite volume, phase transitions are usually turned into smooth crossovers. This is not the case for the Casimir ``plate-plate'' geometries where, apart from the case of (1+1) dimensions, at least one of the space directions is unbounded.

\subsection{${{\mathbb C}P}^{\,N-1}$ sigma model in (1+1) dimensions}

The (1+1) dimensional ${{\mathbb C}P}^{\,N-1}$ sigma model is often considered as a toy model which mimics all essential nonperturbative features of QCD~\cite{Novikov:1984ac}: both models exhibit the asymptotic freedom and dynamically generate a mass gap, they have nonperturbative condensates and topological defects. Similarly to QCD, the ${{\mathbb C}P}^{\,N-1}$ sigma model has two phases at a finite temperature: a confining (``Coulomb'') low-temperature phase and a deconfining (``Higgs'') high-temperature phase. 

The classical action of the 1+1 dimensional ${{\mathbb C}P}^{\,N-1}$ sigma model 
\beqn
S = \int d t \int d x\left[ (D_\mu n_i)^* (D^\mu n_i) - \lambda (n_i^* n_i - r) \right], \qquad r = \frac{4 \pi}{g^2},
\label{eq:S:CP}
\eeqn
describes the dynamics of $N$ complex scalar fields $n_i$ (with $i = 1, \dots, N$) coupled via the covariant derivative $D_\mu = \partial_\mu - i g A_\mu$ to the non-propagating Abelian gauge field $A_\mu$. The model possesses the local Maxwellian $U(1)$ symmetry $n_i(x) \to e^{i \omega(x)} n_i(x)$ and $A_\mu(x) \to A_\mu(x) + \partial_\mu \omega(x)$.

The Lagrange-multiplier field $\lambda = \lambda(t,x)$ imposes the constraint $n_i^* n_i = r$ which determines the ``length'' $r$ of the complex vector $(n_1, \dots n_N)$ via the coupling $g$ in Eq.~\eq{eq:S:CP}.  In an unbounded space, a one-loop analysis shows that the renormalized coupling runs with the momentum scale~$\mu$:
\beqn
r(\mu) \equiv \frac{4 \pi}{g^2(\mu)} = \frac{N}{2 \pi} \log \frac{\mu}{\Lambda}, \qquad \lambda = \Lambda^2,
\label{eq:Lambda}
\eeqn
where $\Lambda \equiv m$ is the dynamically generated mass. 

What happens with the confining vacuum of the ${{\mathbb C}P}^{\,N-1}$ model in-between the ``plates'' and what is the associated Casimir energy? Since in one spatial dimension the ``plates'' are isolated points, the problem should be formulated on a finite spatial interval $ -L/2 \leq x \leq L/2$ with certain boundary conditions at its ends $x = \pm L/2$. There are three series of papers devoted to this problem in the current literature: \cite{ref:Konishi:1,ref:Konishi:2,ref:Konishi:3}, \cite{ref:Nitta:1}, and \cite{ref:Milekhin:1,ref:Milekhin:2}. We will briefly review them below.

It is convenient first to assume that the condensate of the field $n_i$ develops only in one ``classical'' $i=1$ component, $\avr{n} = \sigma \delta_{n,1}$, while $\avr{n_i} = 0$ for 
the ``quantum'' part with $i = 2, \dots, N$. Integrating out the latter, one gets the effective action for the condensates $\sigma$ and~$\lambda$:
\beqn
S_{\mathrm{eff}} = \int d t \int d x\left[ (N-1) \log \left(- D^*_\mu D^\mu + \lambda \right) 
+ (D_\mu \sigma)^* (D^\mu \sigma) - \lambda (|\sigma|^2 - r)  \right].
\label{eq:S:eff}
\eeqn
Working in the large-$N$ approximation, turning the condensate $\sigma$ to the real axis and ignoring the non-dynamical gauge field ($A_\mu = 0$), one gets total energy of the system~\cite{ref:Konishi:1}:
\beqn
E = N \sum_n \omega_n + \int_0^L d x \,\left[ (\partial_x \sigma)^2 + \lambda (\sigma^2 - r)\right].
\label{eq:E:sigma}
\eeqn
The first term corresponds to the Casimir sum over the eigenenergies $\omega_n$ of quantum fluctuations of the $n_i$ fields ($i = N-1$) on top of the classical background of the condensate $\lambda(x)$:
\beqn
\left[ -\partial_x^2 + \lambda(x) \right] f_n(x) = \omega^2 f_n(x).
\label{eq:spectrum}
\eeqn
The second term in Eq.~\eq{eq:E:sigma} is the classical energy of the condensate $\sigma(x)$ itself. The condensate obeys the extremization equation determined by the variation of the action~\eq{eq:S:eff} with respect to  $\sigma$:
\beqn
\partial_x^2 \sigma(x) - \lambda(x) \sigma(x) = 0.
\label{eq:ground:state}
\eeqn
Both condensates $\lambda$ and $\sigma$ may depend on the coordinate $x$ because on a finite interval the translational invariance is lost. This spatial dependence makes the problem very difficult as the ground state is determined by a global minimum of the energy~\eq{eq:E:sigma}, which involves the Casimir part. The latter contribution involves the full spectrum of quantum fluctuations $\omega_n$ which, in turn, depends on the classical condensates in a nonlocal way via equation~\eq{eq:spectrum}.

In an unbounded space $(L \to \infty)$ the translational invariance is restored and the condensates $\sigma$ and $\lambda$ should become space-independent quantities (unless the translational symmetry is spontaneously broken; see below). Then Eq.~\eq{eq:ground:state} implies that in the ground state $\sigma \lambda = 0$, so that at least one of these condensates must vanish. Therefore, one should distinguish the following two phases:

\begin{itemize}

\item[$\blacktriangleright$] the ``confinement'' phase, characterized by the dynamically generated mass $\sqrt{\lambda} \equiv \Lambda \neq 0$ \eq{eq:Lambda} and the vanishing $n$-field condensate $\sigma = 0$. Due to the presence of the mass gap $\Lambda$ the energy density per unit length of the system is a quantity the order of $\Lambda^2$ so it grows with the linear length of the system in analogy with a confining string in QCD.

\item[$\blacktriangleright$] the ``deconfinement'' (Higgs) phase with a vanishing mass $\lambda = 0$ and the condensed field $\sigma\neq 0$. The ``string tension'' is absent while the condensate breaks the $U(1)$ symmetry.
\end{itemize}

According to Eq.~\eq{eq:ground:state}, on a finite interval the system may also develop a third phase, in which both condensates are nonzero, $\lambda(x) \sigma(x) = \partial_x^2 \sigma(x) \neq 0$ if the condensate $\sigma(x)$ is not a constant:
\begin{itemize}
\item[$\blacktriangleright$] the ``mixed'' phase possesses both the inhomogeneous dynamical mass $\sqrt{\lambda} \neq 0$ (as in the confining phase), and the inhomogeneous condensate $\sigma \neq 0$ (as in the non-confining phase).
\end{itemize}

Properties of a field theory in a finite geometry depend crucially on conditions applied to the fields at the boundary of the system. For example, a free massless scalar field on a finite one-dimensional interval has a negative Casimir energy if the field is subjected either to the Dirichlet-Dirichlet (D-D) or to Dirichlet-Dirichlet (N-N) boundary conditions, applied at the same time at both ends of the interval. If one associates the ends with impurities, then these impurities would tend to attract each other. If the boundary conditions are of a mixed type (D-N) then the Casimir energy is a positive quantity and the impurities would repel each other.

In the ${{\mathbb C}P}^{\,N-1}$ sigma model it is natural to set boundary conditions on the $n_i$ fields only. Indeed, the fields $\lambda$ and $A_\mu$ are not constrained at the boundaries because the Lagrange multiplier $\lambda$ should enforce the classical condition condition $n_i^*n_i = r$ both in the bulk and at the boundary while the gauge field $A_\mu$ is not a propagating degree of freedom at all.

Since the matter field $n_i$ is a multicomponent field, we have various choices for boundary conditions as we may imply different (Dirichlet or Neumann) conditions for different combinations of the components ($i, j, k=1, \dots N$) at the opposite ends of the interval ($x = \pm L/2$):
\beqn
\mbox{D${}_r$:} \quad n_i (x) = \sqrt{r}; 
\qquad
\mbox{D${}_0$:} \quad n_j (x) = 0; 
\qquad
\mbox{N:} \quad D_x n_k (x) = 0.
\label{eq:BCs}
\eeqn

\begin{figure}[!thb]
\begin{center}
\includegraphics[scale=0.4,clip=true]{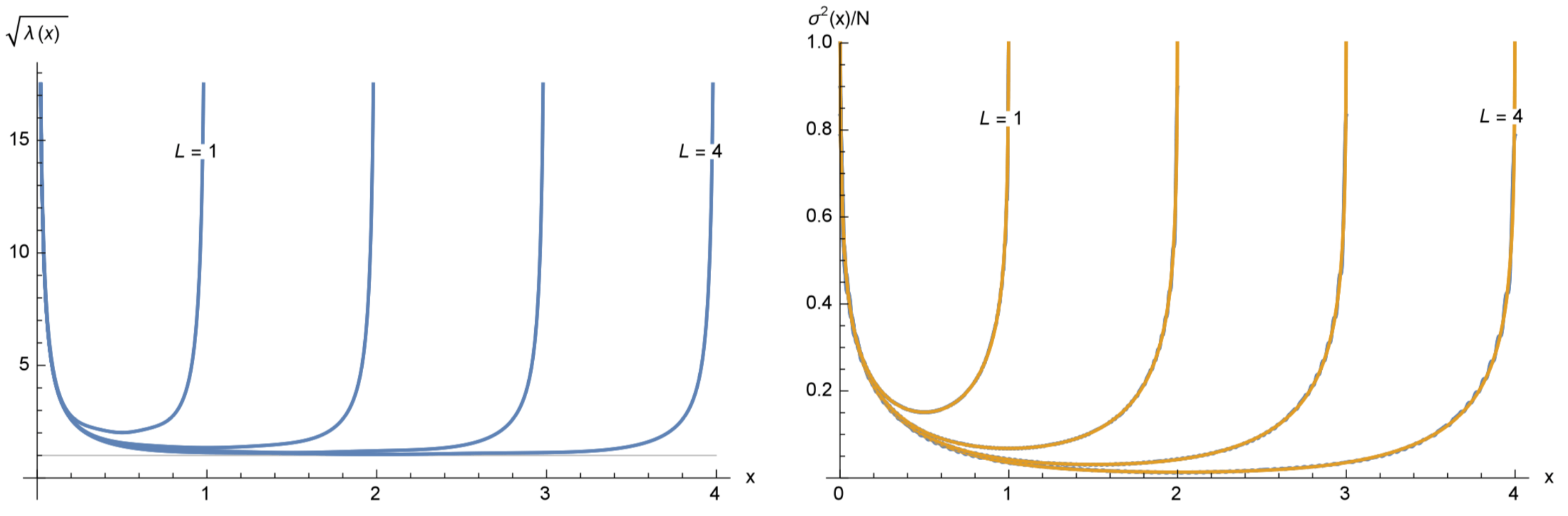} 
\end{center}
\caption{
The mass gap $\sqrt{\lambda(x)}$ and the matter-field condensate (squared and normalized), $\sigma^2(x)/N$, as functions of the coordinate $x$ for various lengths $L$ of the interval (here $0 < x < L$), in units of the mass gap $\Lambda$ of the unbounded ($L\to \infty$) system. Both plots apply to the D-D and to N-N boundaries. (From Ref.~\cite{ref:Konishi:1}).}
\label{fig:condensates}
\end{figure}

In Ref.~\cite{ref:Konishi:1} the large-$N$ ${{\mathbb C}P}^{\,N-1}$ sigma model has been studied on a finite interval with the D-D boundary conditions [so that the D${}_r$ (D${}_0$) requirement is applied to the $i=1$ ($j=2,\dots,N$) component(s) at the both ends of the interval] and with the N-N boundary conditions~\eq{eq:BCs}. In both cases the model was found to possess the unique ``mixed'' phase where both the mass gap $m = \sqrt{\lambda}$ and the condensate $\sigma$ are non-zero. These quantities turned out to be functions of the coordinate $x$ which diverge at the boundaries of the system at $x \to \pm L/2$. The divergence is an expected phenomenon since the coupling $g$, as well as the length of the field $r = n_i^* n_i$, gets renormalized $r = r(\mu)$ at the boundary according to Eq.~\eq{eq:Lambda}. The renormalization scale is governed by the distance to the boundary, $\mu = 1/x$, while the length scale $\Lambda$ is determined by the dynamically generated mass in the unbounded theory. The numerical solutions for the condensates, obtained in Ref.~\cite{ref:Konishi:1} and shown in Fig.~\ref{fig:condensates}, are perfectly consistent with this behavior. 

As the length of the system increases, $L \to \infty$, the mass gap approaches its infinite-volume value $\sqrt{\lambda} \to \Lambda = \mathrm{const}$ (shown by a thin line in the left panel of Fig.~\ref{fig:condensates}) while the scalar condensate $\sigma$ vanishes. Thus, the mixed  phase smoothly approaches the confining phase in the large volume limit. No phase transition is observed between these phases. The total Casimir energy of the system gently interpolates between a standard short-distance behavior of $O(N)$ free (in a leading order) complex scalar fields and the long-distance non-perturbative string-like behavior with the string tension determined by the dynamically generated mass scale $\Lambda$~\cite{ref:Konishi:2}:
\beqn
E(L) = \left\{ 
\begin{array}{ll}
- \frac{N \pi}{6 L}, \qquad & L \ll 1/\Lambda, \\[1mm]
\frac{N \Lambda^2  L}{4 \pi}, \qquad & L \gg 1/\Lambda.
\end{array}
\right.
\eeqn
A similar situation holds for generic Dirichlet conditions, in which the directions of the condensates $\avr{n_i}$ at the opposite boundaries~\eq{eq:BCs} are misaligned. The Casimir energy density is a monotonically increasing function of a degree of misalignment, which takes its global minimum when the boundary condensates are parallel in the ${{\mathbb C}P}^{\,N-1}$ space~\cite{ref:Konishi:3}.

In an alternative, analytical approach of Ref.~\cite{ref:Nitta:1} the authors provide a support for the existence of two phases in the same model on an interval with the D-D boundary conditions. The confining (non-confining) phase was argued to be realized at large (small) length of the interval. The Casimir force turned out to be repulsive in the Higgs phase, while in the confining phase the force may change from the repulsive (small $L$) to an attractive (large $L$) regime. The ``two-phases'' ground state proposed in Ref.~\cite{ref:Nitta:1} differs from the ``one-phase'' ground state found in Refs.~\cite{ref:Konishi:1,ref:Konishi:2,ref:Konishi:3} despite both these states are inhomogeneous. In particular, the behavior of the corresponding solutions near the boundaries differ from each other. 

Contrary to the solutions of Ref.~\cite{ref:Konishi:1,ref:Konishi:2,ref:Konishi:3}, the result of Ref.~\cite{ref:Nitta:1} does not have an explicit relation to the dynamical mass scale $\Lambda$ of the homogeneous vacuum in the infinite-volume limit. On the other hand, the homogeneous nature of the true $L \to \infty$ ground state itself was very recently called for re-consideration in Ref.~\cite{Gorsky:2018lnd} in view of the newly found soliton-like inhomogeneous solution~\cite{Nitta:2017uog}. It was pointed out in Ref.~\cite{Gorsky:2018lnd} that  in an infinite volume the inhomogeneous vacuum state may possess lower energy compared to the well-known homogeneous vacuum.

Coming back to the finite-length systems, we notice that homogeneous condensate solutions were shown in Ref.~\cite{ref:Konishi:1} to be incompatible with the mass gap equations on a finite interval with the D-D and N-N boundary conditions~\eq{eq:BCs}. For the D-D conditions the same conclusion has been independently made in Ref.~\cite{ref:Nitta:1}. A constant solution with the D-D boundary conditions would then lead to the existence of the two phases: the deconfinement (Higgs) phase at small $L$ and the confinement phase at large $L$, while favoring a unique phase -- albeit translationally invariant -- with the N-N boundaries for all components of the fields~\cite{ref:Milekhin:1}. 

It was argued in Ref.~\cite{ref:Milekhin:2} that the constant condensates are compatible with the mixed boundary conditions in a ${{\mathbb C}P}^{\,2N}$ model, where $N$ components ($i = 1, \dots, N$) of the field $n_i$ obey mixed Dirichlet-Neumann conditions [more precisely, D${}_0$-N according to Eq.~\eq{eq:BCs}], another $N$ components ($i = N+1, \dots, 2N$) satisfy the ``mirrored'' N-D${}_0$ conditions while the remaining single $(2N+1)$'th component satisfies the N-N conditions at the both ends. In this specific case the model should indeed possess two translationally-invariant phases separated by a phase transition at the critical distance $L_c = \pi/(4 \Lambda)$~\cite{ref:Milekhin:2}. If one chooses instead the D${}_0$-D${}_0$ conditions for $N$ components and N-N conditions for another set of $N$ components, then the ground state may be chosen in a homogeneous form, the finite-$L$ phase transition does not exist, and the model contains only one (confinement) phase~\cite{ref:Milekhin:2}. Thus, the boundary conditions substantially affect the ground state and the phase structure of the ${{\mathbb C}P}^{\,N-1}$ sigma model on a finite interval.

\subsection{Compact electrodynamics in (2+1) dimensions}

The compact electrodynamics (or ``compact QED'') is another toy model which has interesting nonperturbative features similar to those of QCD: the linear confinement of electric charges, the mass gap generation and the presence of topological defects in physically relevant cases of two and three spatial dimensions. In the case of two spatial dimensions these phenomena may be treated, in a weak coupling regime, using analytical techniques~\cite{Polyakov:1976fu}. Apart from the role of the toy model used to mimic the mentioned phenomena in particle physics, the compact QED also serves as an effective macroscopic model in a wide class of condensed matter systems~\cite{ref:book:Herbut,ref:book:Kleinert}. Below we will briefly describe properties of the Casimir effect in the compact QED obtained in first-principles simulations of lattice field theory (earlier lattice numerical calculations in non-interacting regimes of similar models were done in Ref.~\cite{ref:Oleg}).

In the 3d Euclidean formulation, the (2+1)d compact QED has the following lattice action
\beqn
S[\theta] = \beta \sum_x \sum_{\stackrel{\mu,\nu =1}{{}^{\mu < \nu}}}^3 \left(1 - \cos \theta_{P_{\mu\nu}} \right)\,,
\qquad 
\beta = \frac{1}{g^2 a}\,,
\label{eq:S}
\eeqn
where the sum is taken over all elementary plaquettes $P_{x,\mu\nu}$ of the lattice. The lattice gauge field $\theta_{x,\mu} \in [-\pi,+\pi)$ is a compact Abelian variable (hence the name ``the compact QED'') defined at each link $l_{x,\mu}$ of the lattice. They enter the action~\eq{eq:S} via the plaquette angles $\theta_{P_{x,\mu\nu}} = \theta_{x,\mu} + \theta_{x+\hat\mu,\nu} - \theta_{x+\hat\nu,\mu} - \theta_{x,\nu}$, which play the role of the lattice field strength. The lattice coupling constant $\beta$ is related to the lattice spacing (the length of an elementary link) $a$ and to the electric charge $g$. Notice that the electric charge is a dimensionful quantity $[g] = {\text{mass}}^{1/2}$ in (2+1) dimensions.

In the continuum limit ($a\to 0$) the lattice plaquette variable tends, for small fluctuations of the lattice photon fields, to its continuum version $\theta_{P_{x,\mu\nu}} = a^2 F_{\mu\nu}(x) + O(a^4)$ with $F_{\mu\nu} = \partial_\mu A_\nu - \partial_\nu A_\mu$. Consequently, the lattice action~\eq{eq:S} becomes the standard photon action.

The compact QED~\eq{eq:S} possesses also the monopole singularities with the density
\beqn
\rho_x = \frac{1}{2\pi} \sum_{P \in \partial C_x} {\bar \theta}_P \, \in \Z\,,
\label{eq:rho:lattice}
\qquad
{\bar \theta}_P = \theta_P + 2 \pi k_P \in [-\pi,\pi), \qquad k_P \in \Z,
\label{eq:bar:theta}
\eeqn
where the integer number $k_P$ is chosen in such a way that the physical plaquette angle ${\bar \theta}_P$ is limited to the interval $[-\pi,\pi)$. The sum in Eq.~\eq{eq:rho:lattice} goes over all faces $P$ of an elementary cube $C_x$. In two spatial dimensions, the monopole is an instanton-like topological object which appears due to the compactness of the gauge group. The compactness comes from the invariance of the action \eq{eq:S} under the discrete transformations of the lattice field strengths: $\theta_P \to \theta_P + 2 \pi n$ with $n \in \Z$.

Thus, the model~\eq{eq:S} describes the dynamics of photons (weak fields) and monopoles (strong fields). The photons characterize a perturbative regime relevant, in particular, to a short-distance Coulomb potential between test electric charges. The monopole dynamics is responsible for nonperturbative effects such as the long-range linear potential between the oppositely charged particles:
\beqn
V(L) = \sigma L\,, 
\qquad
\sigma = \frac{4 g \sqrt{\varrho}}{\pi}\,, 
\qquad 
\varrho \equiv \avr{|\rho_x|}
\label{eq:V:L}
\eeqn
where the quantity $\sigma$ -- given in Eq.~\eq{eq:V:L} in the dilute gas approximation~\cite{Polyakov:1976fu} -- is interpreted as a tension of a string which spans between static particle and antiparticle, and confines them into a chargeless bound state. The presence of monopoles generates the mass gap
\beqn
m = \frac{2\pi \sqrt{\rho}}{g}\,,
\label{eq:mass:gap}
\eeqn
and drives a finite-temperature phase transition at certain critical temperature $T = T_c$.

In two spatial dimensions the standard Casimir problem is formulated for one-dimensional objects (``wires''). A static and infinitely thin wire, made of a perfect metal, forces the tangential component of the electric field $\vec E$ to vanish at every point $\bx$ of the wire, $E_\| (\bx) = 0$. The wire does not affect the pseudoscalar magnetic field $B$. In a covariant form the corresponding boundary conditions are as follows:
\beqn
F_{\mu\nu}(x) s^{\mu\nu}(x) = 0\,, 
\qquad
s^{\mu\nu}(x) = \int d^2 \tau \, \frac{\partial {\bar x}^{[\mu,}}{\partial \tau_1} \frac{\partial {\bar x}^{\nu]}}{\partial \tau_2} \,
\delta^{(3)}\bigl(x - {\bar x}(\vec\tau)\bigr), \quad
\label{eq:Casimir:F}
\eeqn
where $F_{\mu\nu} = \partial_{[\mu,} A_{\nu]} \equiv \partial_\mu A_\nu - \partial_\nu A_\mu$ is the field strength tensor and  $s_{\mu\nu}$ is the local surface element of the world sheet of the wire described by the vector function ${\bar x}_\mu = {\bar x}_\mu({\vec \tau})$ and parametrized by the two-vector $\vec\tau = (\tau_1,\tau_2)$.

\begin{figure}[!thb]
\begin{center}
\begin{tabular}{ccc}
\includegraphics[scale=0.065,clip=true]{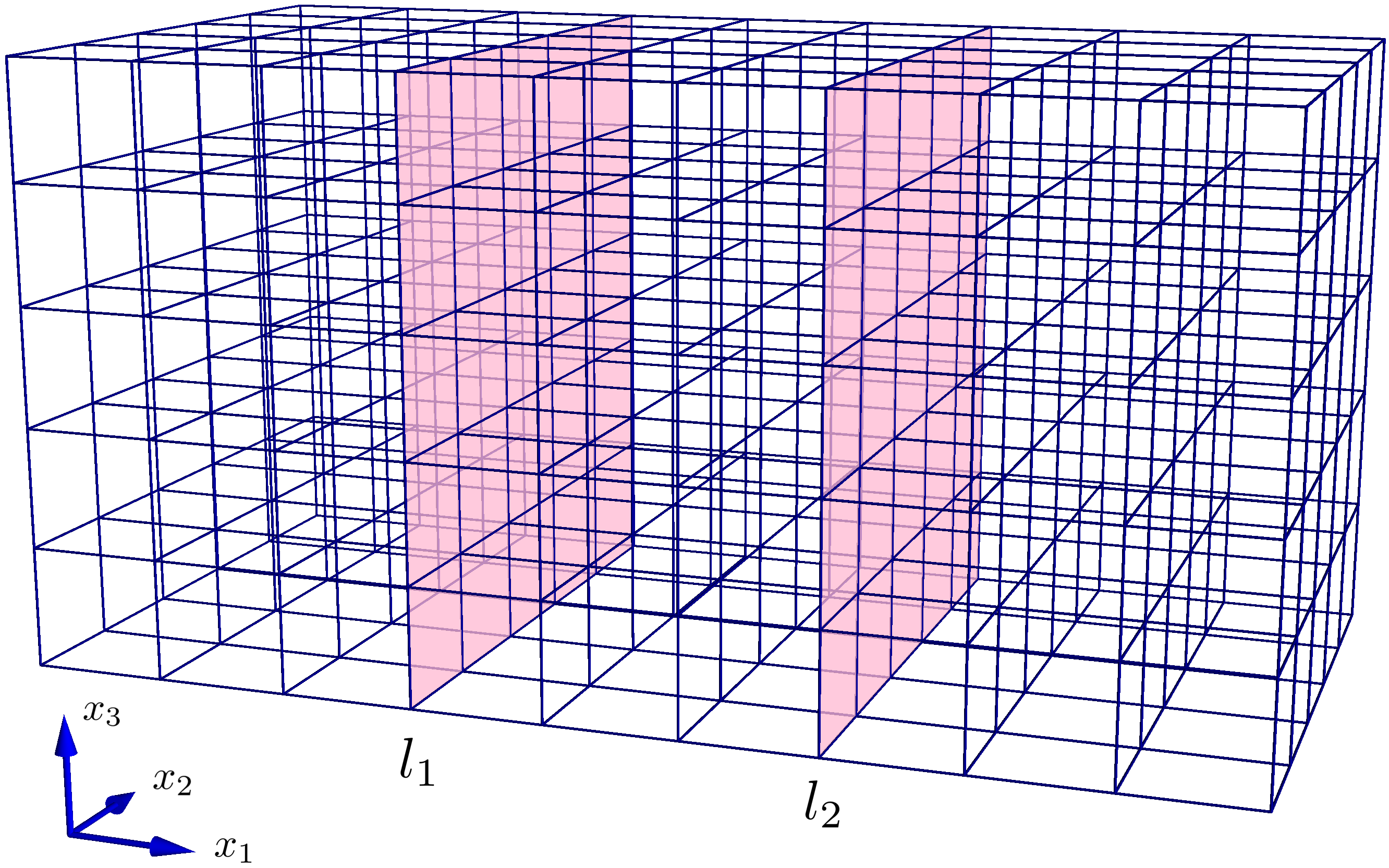} & 
\includegraphics[scale=0.28,clip=true]{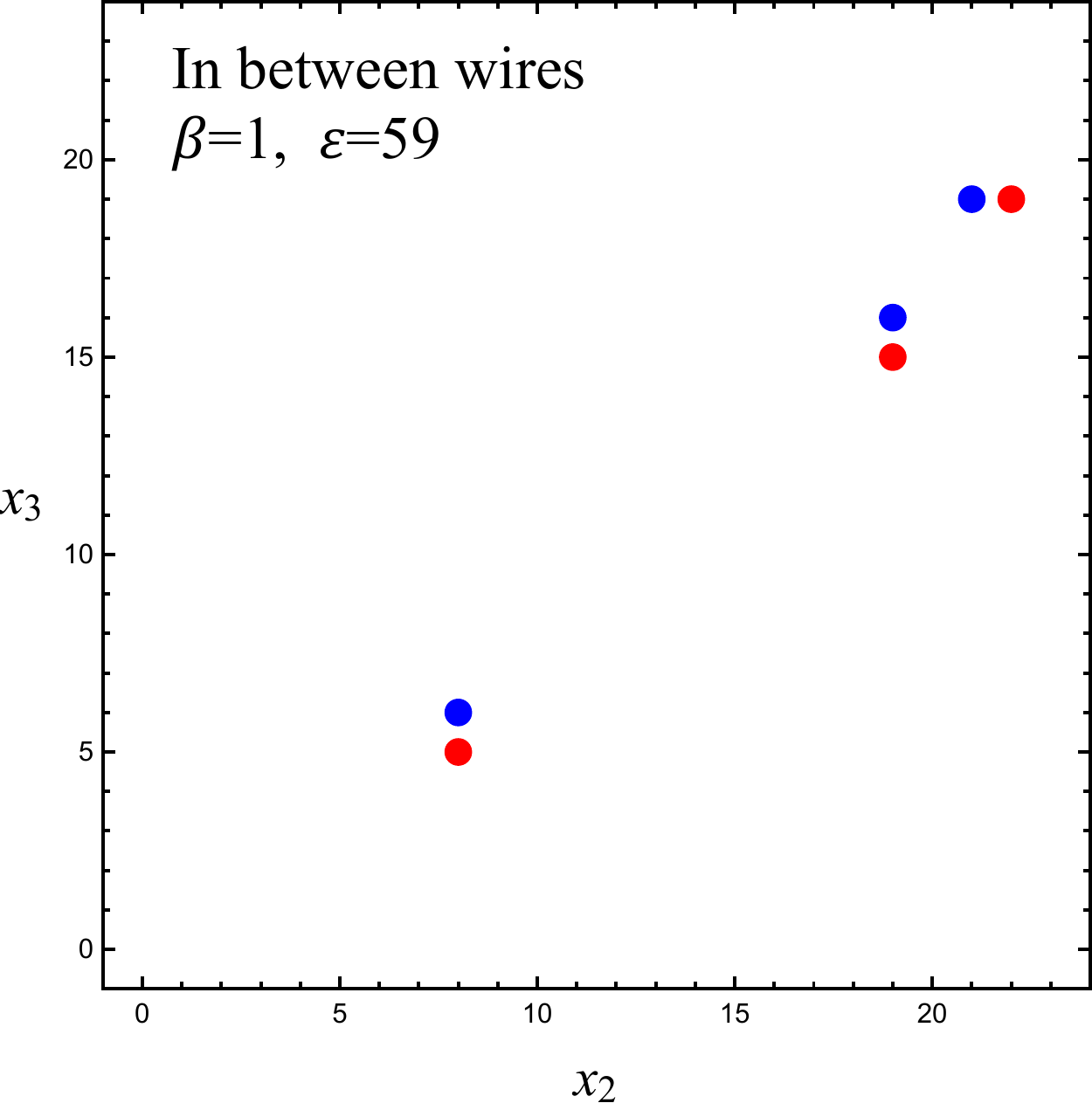} & 
\includegraphics[scale=0.28,clip=true]{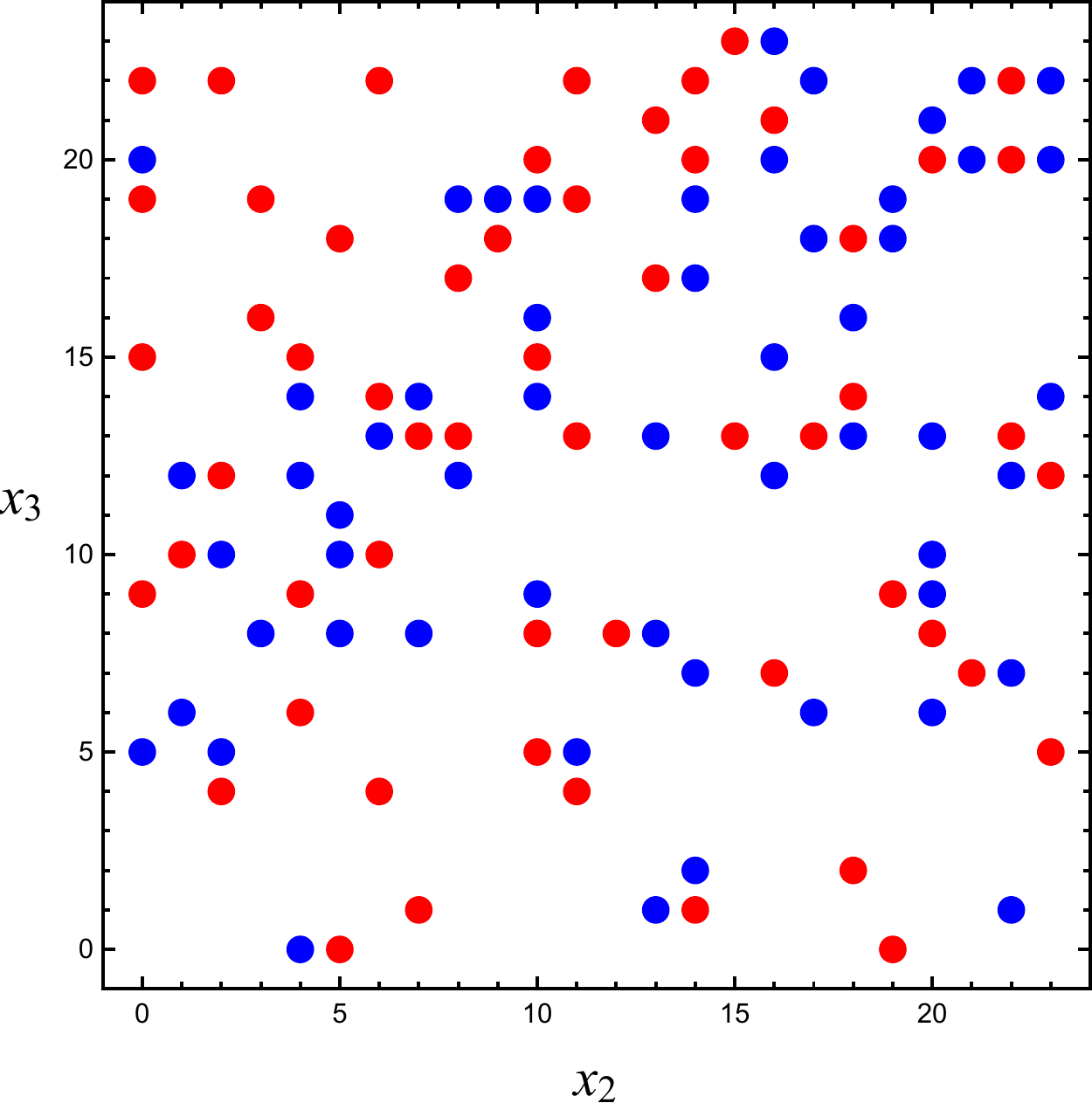} \\
(a) & (b) & (c)
\end{tabular}
\end{center}
\caption{(a) The geometry of the Casimir problem in (2+1)d lattice: the shadowed planes (``plates'') indicate the plaquettes of the wire world-surfaces where the boundary condition~\eq{eq:F01:latt:3D} is implemented (from \cite{Chernodub:2016owp}).
Examples of typical configurations of monopoles (blue) and antimonopoles (red) in (b) a slice in between closely spaced plates and (c) in a space outside the plates
(from \cite{Chernodub:2017mhi}).}
\label{fig:geometry:plane}
\end{figure}

It is convenient to consider two static straight wires directed, for example, along the $x_2$ axis and separated along the $x_1$ direction at $x_1 = l_1$ and $x_1 = l_2$ as shown in Fig.~\ref{fig:geometry:plane}(a). The $x_3$ axis is associated with the Euclidean ``time'' direction. A lattice analogue of the Casimir boundary condition~\eq{eq:Casimir:F},
\beqn
\cos\theta_{x,23} {\biggl|}_{x_1 = l_a} = 1, \qquad a = 1,2\,,
\label{eq:F01:latt:3D}
\eeqn
ensures that at the world-surfaces of the wires the lattice field strength vanishes. A simplest way to implement the boundary condition~\eq{eq:F01:latt:3D} is to add a set of Lagrange multipliers to the standard Abelian action~\eq{eq:S} via the plaquette-dependent gauge coupling:
\beqn
S_{\varepsilon}[\theta] = \sum_P \beta_P(\varepsilon) \cos \theta_P\,,
\qquad
\beta_{P_{x,\mu\nu}} (\varepsilon) = \beta \bigl[1 + (\varepsilon - 1) \delta_{\mu,2} \delta_{\nu,3} \left(\delta_{x,l_1} + \delta_{x,l_2}\right)\bigr].
\label{eq:beta:P:3d}
\eeqn
The coupling is a function of the dielectric permittivity $\varepsilon$ of the wire. At $\varepsilon = 1$ the wires are absent. In the limit $\varepsilon \to + \infty$ the components of the physical lattice field-strength tensor~\eq{eq:bar:theta} vanish at the world surfaces of the wires as required by Eq.~\eq{eq:F01:latt:3D}.

The Casimir energy density corresponds to a component of the canonical energy-momentum tensor, $T^{00} = (E^2 + B^2)/(2 g^2)$. The numerical calculations reveal that the presence of the Abelian monopoles affects the Casimir effect nonperturbatively~\cite{Chernodub:2017mhi}. At large separations between the wires the Casimir energy becomes screened by the mass gap~\eq{eq:mass:gap}. At small separations, it is the wires that affect the monopoles: as the wires approach each other, the relatively dense monopole gas in between them gets continuously transformed into a dilute gas of monopole-antimonopole pairs, as it is illustrated in Figs.~\ref{fig:geometry:plane}(b) and (c)~\cite{Chernodub:2017gwe}. The geometry-induced binding transition is similar to the infinite-order phase transition of a Berezinskii--Kosterlitz--Thouless (BKT) type~\cite{ref:BKT} which occurs in the same model at a finite-temperature~\cite{ref:binding:cU1}. 

\begin{figure}[!thb]
\begin{center}
\begin{tabular}{cc}
\includegraphics[scale=0.45,clip=true]{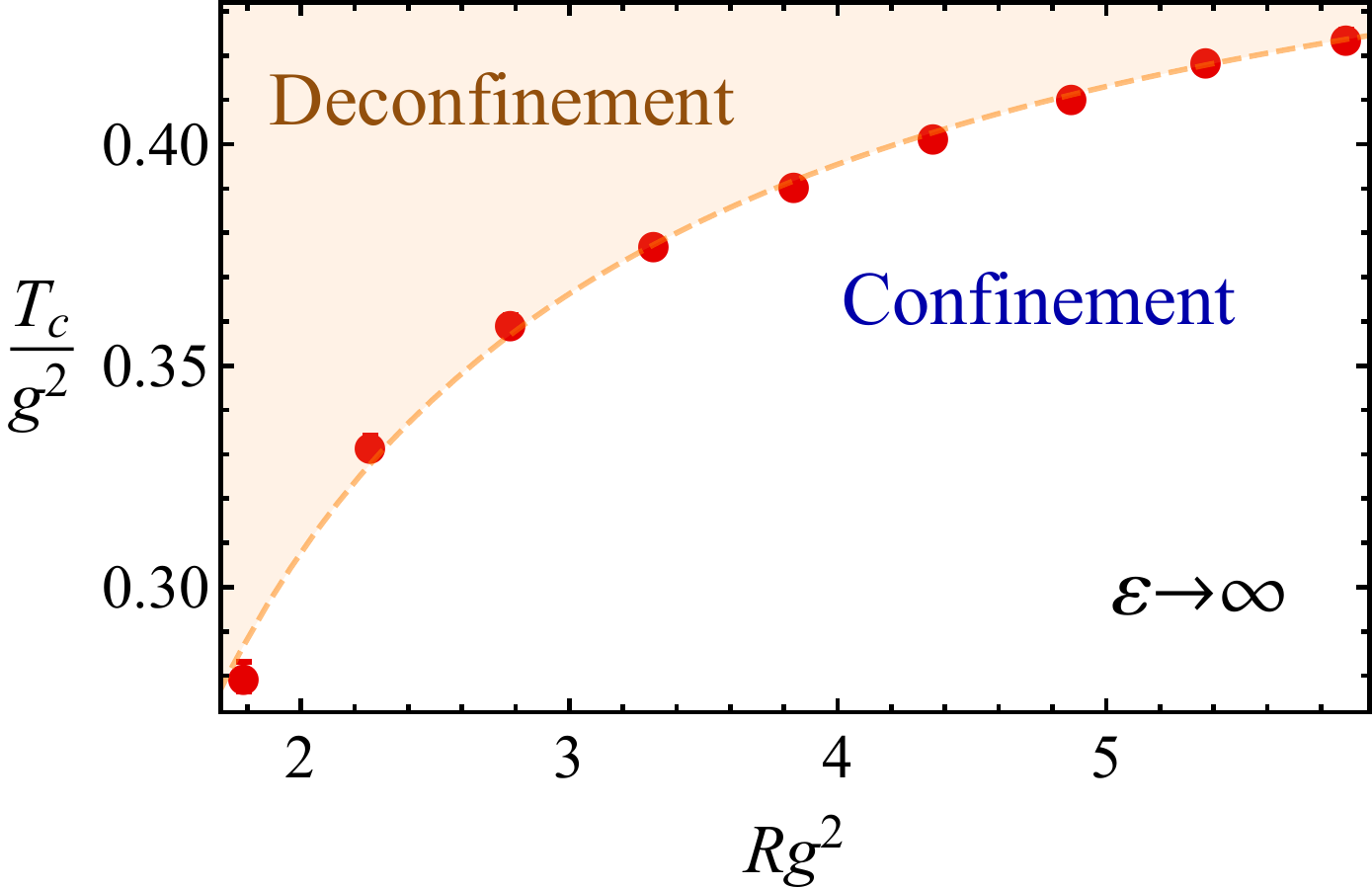} & 
\hskip 10mm
\includegraphics[scale=0.0775,clip=true]{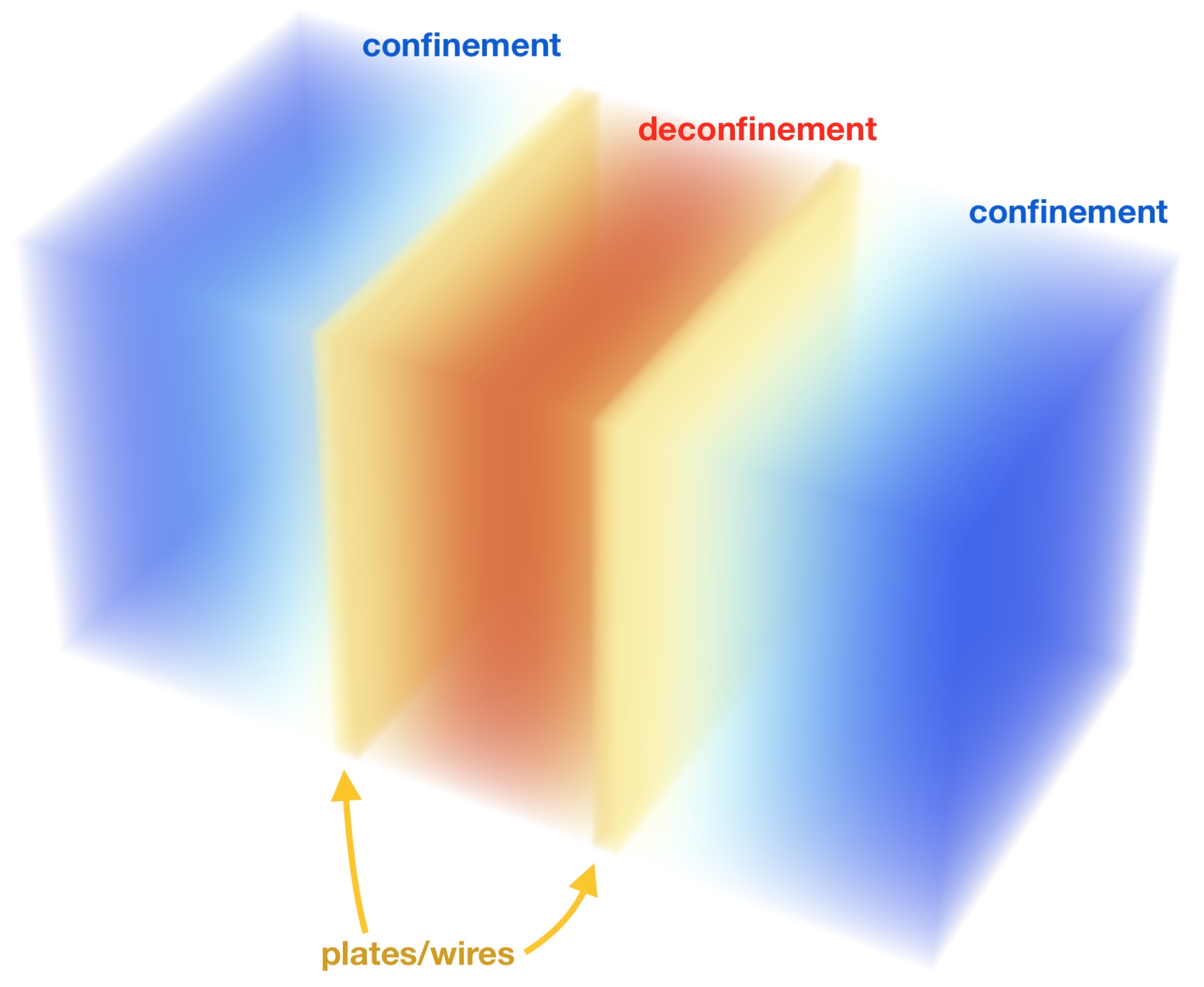} \\
(a) & (b)
\end{tabular}
\end{center}
\vskip -4mm 
\caption{(a) The phase in-between the plates: the critical temperature $T_c$ of the deconfinement transition as the function of the inter-plate distance~$R$ in units of the electric charge $g$ in the ideal-metal limit ($\varepsilon \to \infty$). (b) An illustration of the deconfinement in the space between the plates (from Ref.~\cite{Chernodub:2017gwe}).}
\label{fig:deconfinement}
\end{figure}

The BKT transition is associated with a loss of the confinement property in between the metallic plates because the weak fields of the magnetic dipoles cannot lead to a disorder of the Polyakov-line deconfinement order parameter. This conclusion agrees well expectation with a direct evaluation of the Polyakov line in between the plates~\cite{Chernodub:2017gwe}. Figure~\ref{fig:deconfinement}(a) shows the phase structure of the vacuum of compact electrodynamics in the space between long parallel Casimir wires at finite temperature $T$.  The deconfinement temperature $T_c$ is a monotonically rising function of the interwire distance~$R$. Formally, the charge confinement disappears completely when the separation between the plates becomes smaller than certain critical distance $R = R_c$ determined by the condition $T_c(R_c) = 0$. According to the numerical estimates of Ref.~\cite{Chernodub:2017gwe}, $R_c = 0.72(1)/g^2$\,.

\subsection{Yang-Mills theory in (2+1) dimensions}

The Casimir problem may also be formulated for a non-Abelian gauge theory which possesses inherently nonperturbative vacuum structure. It is instructive to consider a zero-temperature Yang-Mills theory in (2+1) spacetime dimensions with the Lagrangian:
\beqn
{\mathcal L}_{YM} = - \frac{1}{4} F_{\mu\nu}^a F^{\mu\nu, a}, 
\qquad 
F^a_{\mu\nu} = \partial_\mu A_\nu^a - \partial_\nu A_\mu^a + g f^{abc} A_\mu^b A_\nu^c,
\qquad
a = 1, \dots N_c^2 -1,
\label{eq:L:YM}
\eeqn
where $f^{abc}$ are the structure constants of the $SU(N_c)$ gauge group. The model in (2+1) dimensions exhibits both mass gap generation and color confinement similarly to its 3+1 dimensional counterpart. A non-Abelian analogue of the perfect conductor condition~\eq{eq:Casimir:F} is straightforwardly given by a simple substitution of the field strength: $F_{\mu\nu} \to F^a_{\mu\nu}$. 

The theory~\eq{eq:L:YM} can be studied using first-principles numerical lattice simulations with the use of the standard Wilson plaquette action $S_P = \beta_P ( 1 - \frac{1}{2}  {\mathrm{Tr}}\, U_P)$, where the plaquette field strength $U_{P_{x,\mu\nu}} = U_{x,\mu}U_{x+\hat\mu,\nu}U^\dagger_{x+\hat\nu,\mu} U^\dagger_{x,\nu}$ is given by the ordered product of four $SU(2)$ link fields $U_l$ along the plaquette edges. In the bulk, the plaquette lattice couplings $\beta_P \equiv \beta = 4/(a g^2)$ are related to the lattice spacing $a$ similarly to the Abelian case~\eq{eq:beta:P:3d}. At the wire world-surface, shown in Fig.~\ref{fig:geometry:plane}(a), the plaquette couplings are set to infinity. The physical scale for the dimensional couplings is set by the tension $\sigma$ of the confining string in a zero-temperature theory.

\begin{figure}[!thb]
\begin{center}
\begin{tabular}{cc}
\includegraphics[scale=0.054,clip=true]{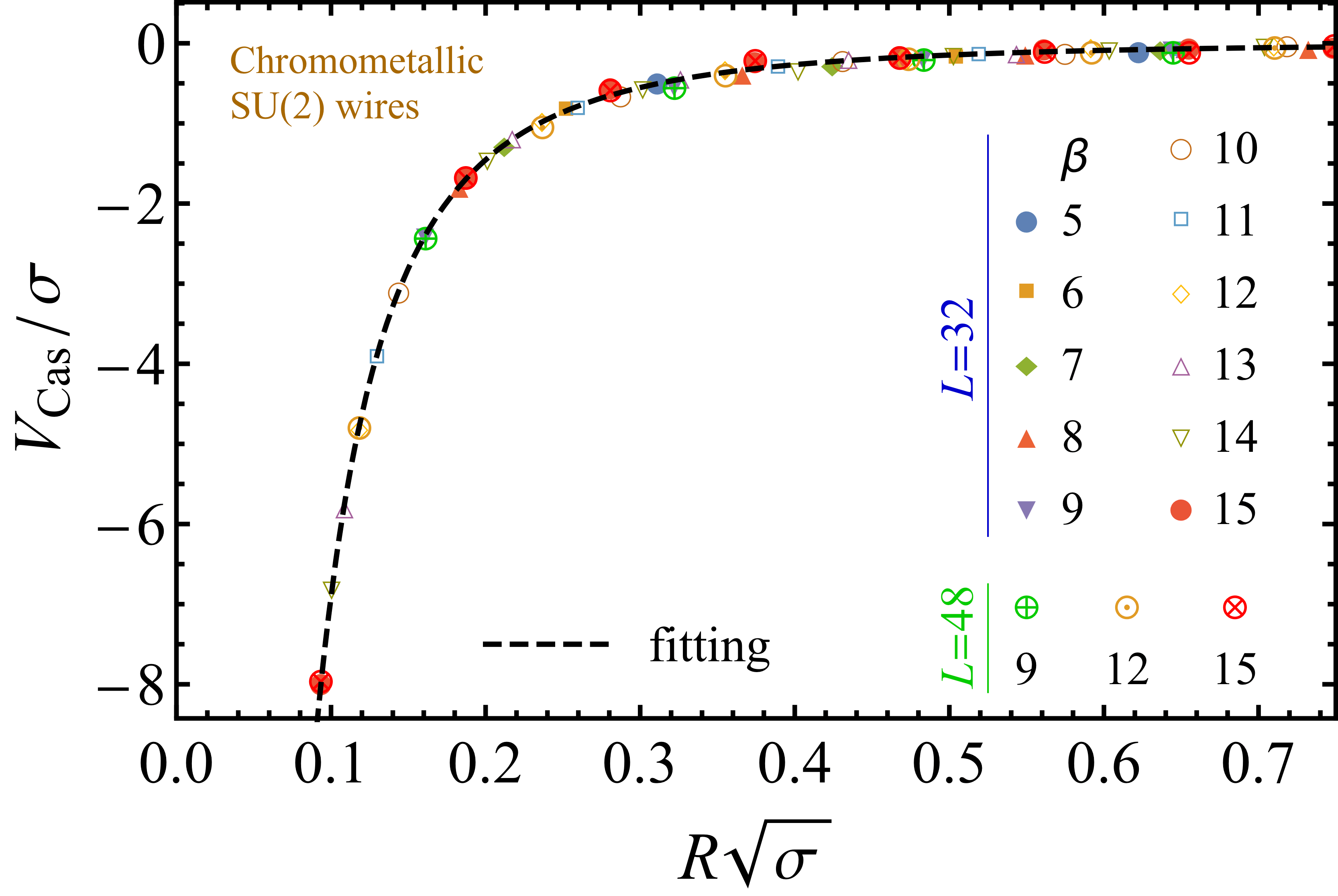} & 
\includegraphics[scale=0.62,clip=true]{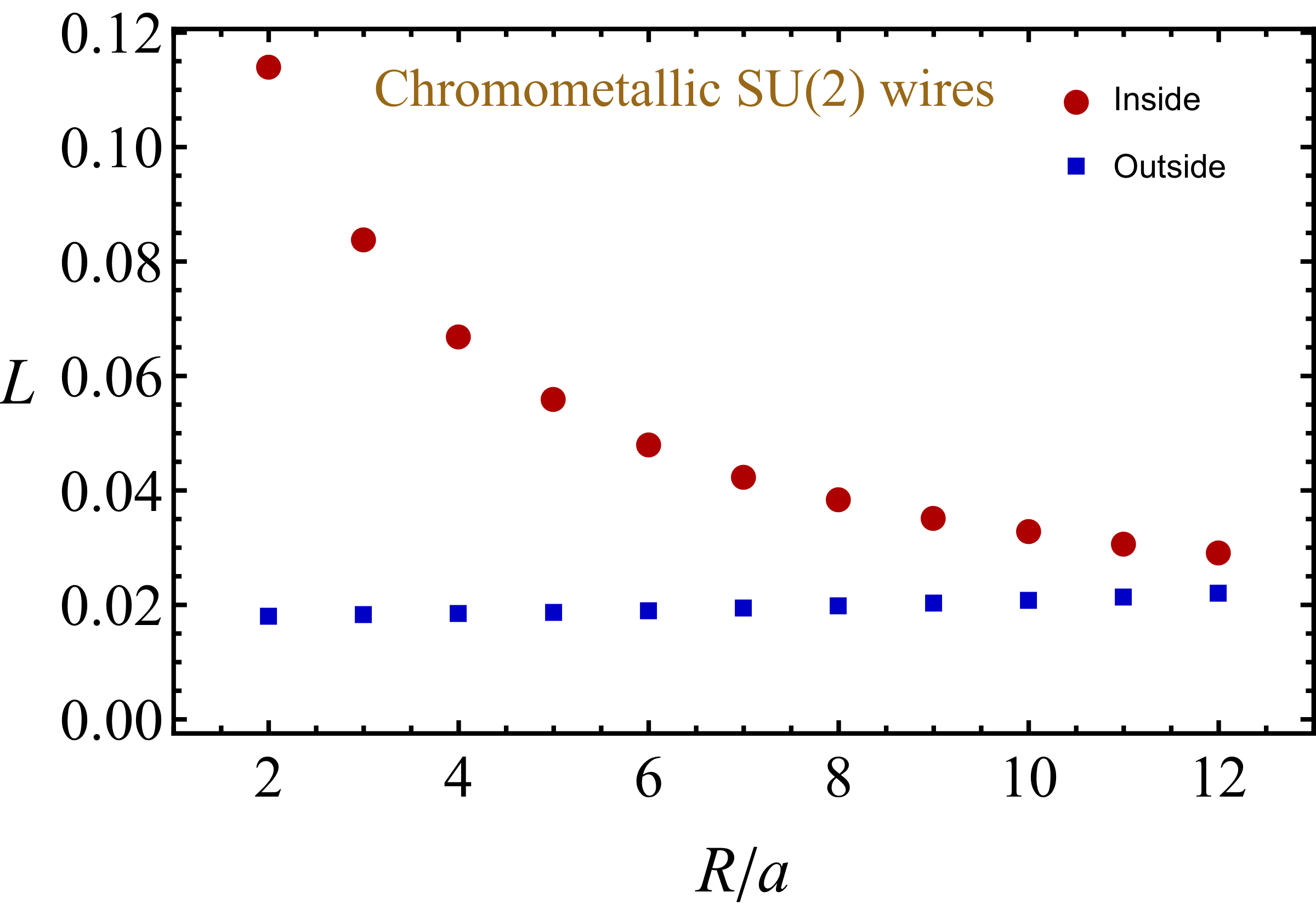} \\
(a) & (b)
\end{tabular}
\end{center}
\vskip -5mm 
\caption{(a) The Casimir potential $V_{\Cas}$ for a chromometallic wire as the function of the distance $R$ between the wires (in units of the string tension~$\sigma$) at various ultraviolet lattice cutoffs controlled by the lattice spacing~$\beta$. The line is the best fit~\eq{eq:V:fit}. (b) A typical expectation value of the absolute value of the mean Polyakov line in the spaces in between and outside the wires vs. the interwire separation~$R$.
(From Ref.~\cite{Chernodub:2018pmt}).}
\label{fig:V}
\end{figure}

The Casimir energy of gluon fluctuations per unit length of the wire is shown in Fig.~\ref{fig:V}(a). The lattice results -- which exhibit the excellent scaling with respect to a variation of the lattice cutoff -- can be described very well by the following function:
\beqn
V_{\Cas}(R) = 3 \frac{\zeta(3)}{16 \pi} \frac{1}{R^2}\frac{1}{(\sqrt{\sigma} R)^{\nu}} e^{- M_{\Cas} R},
\label{eq:V:fit}
\eeqn
where the anomalous power $\nu$ (which controls the short-distance behavior) and the ``Casimir mass'' $M_C$ (which is responsible for the screening at large inter-wire separations) may be determined with the help of a fitting procedure. The values $\nu = 0$ and $M_{\Cas} = 0$ correspond to the Casimir energy of three non-interacting vector particles. In the SU(2) Yang-Mills theory one gets:
\beqn
M_{\Cas} = 1.38(3) \sqrt{\sigma}\,, 
\qquad
\nu_\infty = 0.05(2).
\label{eq:M:infty}
\eeqn
Surprisingly, the Casimir mass $M_{\Cas}$ turns out to be substantially smaller than the mass of the lowest colorless excitation, the $0^{++}$ glueball, $M_{0^{++}} \approx 4.7 \sqrt{\sigma}$  (the latter quantity has been calculated numerically in Ref.~\cite{Teper:1998te}). In Ref.~\cite{Karabali:2018ael} it was shown that the (2+1) Casimir mass may be related to the magnetic gluon mass of the Yang-Mills theory in (3+1) dimensions.

In Fig.~\ref{fig:V}(b) we show the expectation value of the order parameter of the deconfining transition, the Polyakov loop $\avr{L}$, in the space inside and outside the wires. Similarly to the compact Abelian case, the gluons in between the wires experience a smooth deconfining transition as the wires approach each other, thus confirming the qualitative picture shown in Fig.~\ref{fig:deconfinement}(b).

\clearpage

\section{Casimir effect and chiral symmetry breaking}
\label{eq:chiral}

Another important phenomenon of the strong interactions, the spontaneous chiral symmetry breaking, takes place in the fermionic sector of QCD. Many features of the chiral symmetry breaking may be captured by the Nambu--Jona-Lasinio model~\cite{ref:NJL}. We will consider its (1+1) dimensional version which is known as the chiral Gross-Neveu (GN) model~\cite{ref:GN}. In a massless limit, the Lagrangian of this model
\beqn
\cL = i {\bar \psi} {\slashed \partial} \psi + \frac{g}{2} \left[ \left( {\bar \psi} \psi \right)^2 + \left( {\bar \psi} i \gamma_5 \psi \right)^2 \right],
\label{eq:L:chi:GN}
\eeqn
is invariant under the continuous chiral transformations $\psi \to e^{i \gamma^5 \alpha} \psi$ of the $N$-flavor fermion field~$\psi$. The chiral symmetry is spontaneously broken by the combined scalar-pseudoscalar condensate $\Delta = - (\avr{{\bar \psi} \psi} - i \avr{{\bar \psi} i \gamma_5 \psi})/g$ which emerges spontaneously.

The ground state of the chiral GN model~\eq{eq:L:chi:GN} corresponds to an extremum of an effective action which includes contributions from the condensate $\Delta$ itself and from the quantum fluctuations over it. On general grounds, one could expect that in a spatially bounded system (on an interval with equivalent boundary conditions at the both ends) the bosonic condensate will try to push the boundaries of the system outwards. This repulsive force would compete with an attractive force coming from the usual Casimir effect produced by the quantum fluctuations of fermions. 

\begin{figure}[!thb]
\vskip 1mm
\begin{center}
\begin{tabular}{cc}
\includegraphics[scale=0.335,clip=true]{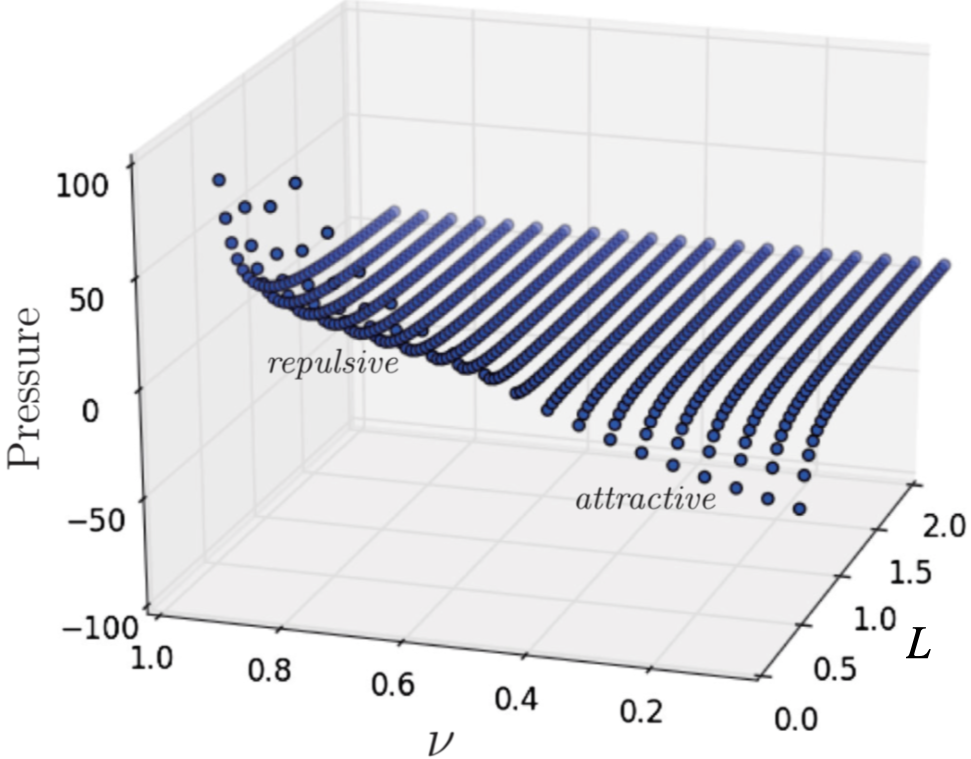} & 
\hskip 10mm
\includegraphics[scale=0.11,clip=true]{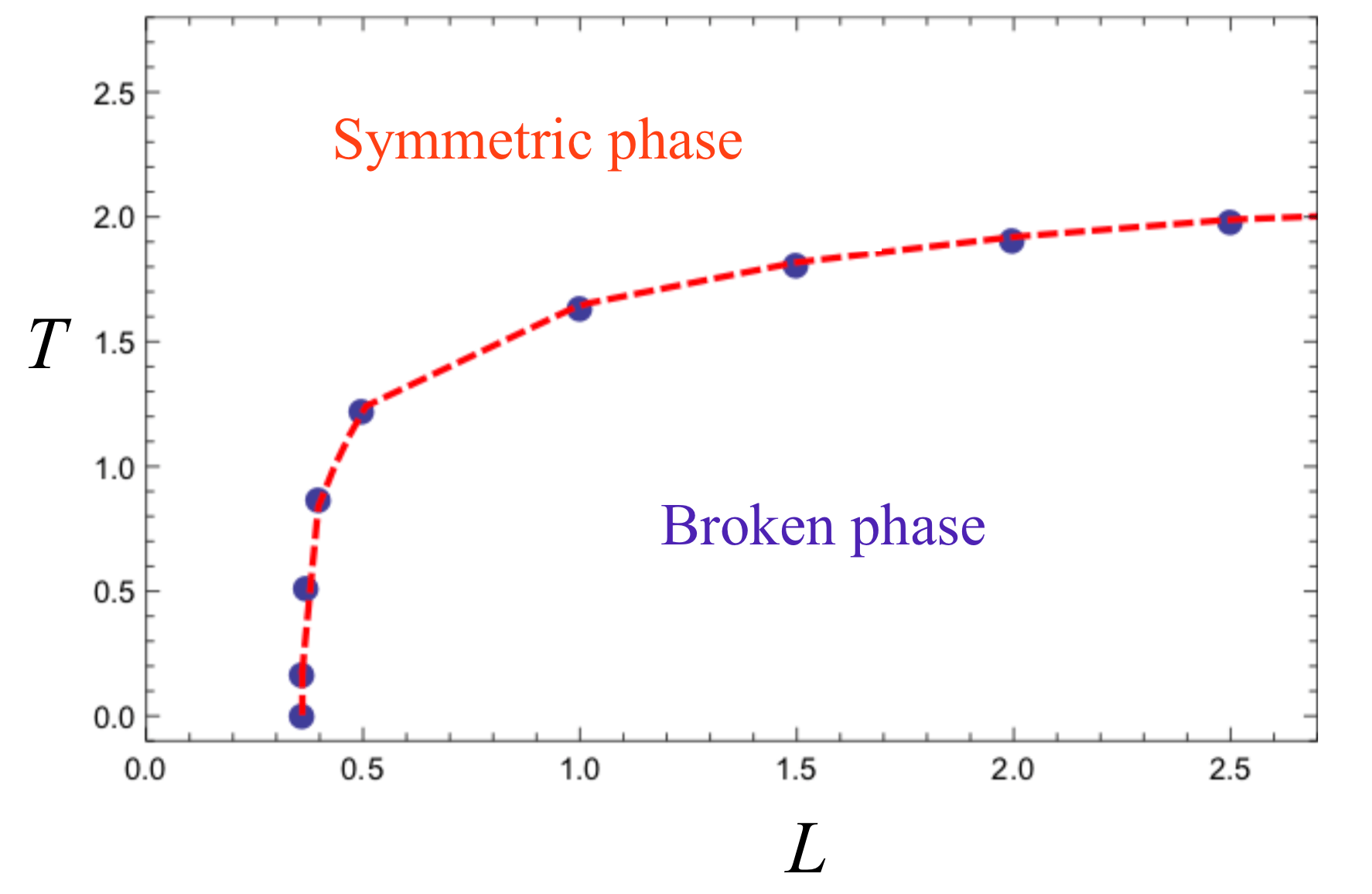} \\
(a) & (b)
\end{tabular}
\end{center}
\vskip -4mm 
\caption{(a) Casimir pressure as a function of the elliptic parameter $\nu$ and the spatial size $L$ of the system in the chiral GN model in (1+1) dimensions~\eq{eq:L:chi:GN}, from Ref.~\cite{Flachi:2017cdo}. (b) The phase diagram of the (3+1) dimensional model of interacting fermions~\eq{eq:L:GN} for the inter-plate distance $L$ and temperature $T$, from Ref.~\cite{Flachi:2013bc}. The dimensional quantities are expressed in units of the coupling $g$.}
\label{fig:fermionic}
\end{figure}

A self-consistent inhomogeneous solution for the ground state of the large-$N$ chiral GN model on an interval was found in Ref.~\cite{Flachi:2017cdo}. The condensate and, consequently, the Casimir energy can be expressed in terms of Jacobi's elliptic functions. The shape of the solution is controlled by the elliptic modulus parameter $\nu$ which, in turn, implicitly depends on the coupling constant $g$ of the model~\eq{eq:L:chi:GN}. The Casimir force exhibits a nontrivial behavior displaying a transition from an attractive to a repulsive regime occurring at a critical value $\nu \approx 0.4$, Fig.~\ref{fig:fermionic}(a). At a small length $L$ of the interval, a decrease (increase) in the coupling $g$, reduces (increases) the elliptic parameter $\nu$ thus leading to an attractive (repulsive) behavior of the Casimir force~\cite{Flachi:2017cdo}.

In (3+1) dimensions, the chiral symmetry breaking in the Casimir double-plate geometry has been studied in the interacting fermionic model with the Lagrangian~\cite{Flachi:2013bc,Flachi:2012pf}:
\beqn
\cL = i {\bar \psi} {\slashed \partial} \psi + \frac{g}{2} \left( {\bar \psi} \psi \right)^2 ,
\label{eq:L:GN}
\eeqn
which is invariant under the discrete $\Z_2$ chiral transformation $\psi \to \gamma^5 \psi$. The theory~\eq{eq:L:GN} is the higher-dimensional analogue of the original Gross-Neveu model~\cite{ref:GN}. 

In an infinite volume the vacuum of this model develops a dynamical chiral condensate $\avr{{\bar \psi} \psi}$ which breaks the $\Z_2$ chiral symmetry. The condensate vanishes and the symmetry is restored at high temperatures via a second order phase transition. In the presence of the approaching Casimir plates, the critical temperature decreases while the phase transition becomes of the first order. As the plates become sufficiently close to each other, the chiral symmetry gets completely restored even at zero temperature. The phase diagram of the interacting fermionic model~\eq{eq:L:GN} in the volume in-between the Casimir plates is shown in Fig.~\ref{fig:fermionic}(b). The restoration of the chiral symmetry due to the shrinking Casimir geometry agree well with the observation that boundary effects restore the chiral symmetry in a chirally broken phase~\cite{Tiburzi:2013vza,Chernodub:2016kxh}.

\section{Conclusions}

We briefly reviewed the Casimir effect in interacting field theories. In the perturbative context of QED, a one-loop correction leads to the expectedly small contribution~\eq{eq:E:Casimir:delta} to the Casimir energy~\cite{Bordag:1983zk}. A two-loop calculation reveals surprising (albeit very tiny) increase of the speed of light in between the plates~\eq{eq:Scharnhorst}. This ``faster-than-light'' phenomenon, known as the Scharnhorst effect~\cite{Scharnhorst:1990sr}, does not lead to causal paradoxes, however~\cite{ref:causality}.

Focusing on non-perturbative features of QCD, we discussed recent studies of the Casimir effect in various field-theoretical models which mimic the physics of confinement,  mass gap generation and/or chiral symmetry breaking.

In (1+1) dimensions, the asymptotic freedom and mass gap generation may be modeled by the ${{\mathbb C}P}^{\,N-1}$ sigma model. The nature of its ground state on a finite spatial interval depends crucially on a precise form of the conditions imposed on the fields at the boundaries~\cite{ref:Konishi:1,ref:Konishi:2,ref:Konishi:3,ref:Nitta:1,ref:Milekhin:1,ref:Milekhin:2}. While the structure of the ground-state condensates is currently under debates (both in infinite space~\cite{Gorsky:2018lnd} and on a finite interval), it is clear that the Casimir force is strongly affected the mass-gap generation phenomenon. 

In (2+1) confining field theories, the Casimir double-wire geometry with a perfect (chromo-) metallic boundary conditions causes a smooth deconfining transition both in the compact electrodynamics~\cite{Chernodub:2016owp,Chernodub:2017mhi,Chernodub:2017gwe} and in Yang-Mills theory~\cite{Chernodub:2018pmt}. Moreover, the Casimir problem in Yang-Mills theory reveals a new intrinsic mass scale which is substantially lower than the lowest $0^{++}$ glueball mass~\cite{Chernodub:2018pmt}. This Casimir mass was argued to be related to the magnetic gluon mass in a finite-temperature Yang-Mills theory in (3+1) dimensions~\cite{Karabali:2018ael}.

In the chiral sector, the interactions may lead to a change of the sign of the Casimir force and affect the pattern of the chiral symmetry breaking in the chiral Gross-Neveu model in (1+1) dimensions~\cite{Flachi:2017cdo}. In a similar model in (3+1) dimensions, the Casimir effect in the double-plate geometry leads to the strengthening of the finite-temperature phase transition and to the decrease of critical temperature associated with chiral symmetry breaking~\cite{Flachi:2013bc}.

\begin{acknowledgments}
The authors thank Michele Caselle, Antonino Flachi, Kenichi Konishi and Muneto Nitta for helpful communications. The research was carried out within the state assignment of the Ministry of Science and Higher Education of Russia (Grant No. 3.6261.2017/8.9). The work was also supported by the project PICS07480 of CNRS (France).
\end{acknowledgments}

\end{document}